\documentclass[reprint,english,twocolumn,prl,showpacs,superscriptaddress,noeprint,floatfix]{revtex4-1}
\usepackage[T1]{fontenc}
\usepackage[latin1]{inputenc}
\usepackage{graphicx}
\usepackage{amssymb}
\usepackage{color}
\usepackage{units}
\usepackage{babel}
\usepackage{longtable}

\providecommand{\tabularnewline}{\\}

\begin{document}

\preprint{LPSC 10-157}

\preprint{KA-TP 36-2010}

\preprint{SMU-HEP 10-xxxxxxxxx }

\title{Nuclear Corrections in Neutrino-Nucleus Deep Inelastic Scattering and their Compatibility with
Global Nuclear Parton-Distribution-Functions Analyses}

\author{\ K.~Kova\v{r}\'{\i}k\,}

\affiliation{LPSC, Universit{é} Joseph Fourier/CNRS-IN2P3/INPG, UMR5821, Grenoble,
F-38026, France}

\affiliation{Karlsruhe Institute of Technology, Karlsruhe, D-76128,Germany}

\author{I.~Schienbein\,}

\affiliation{LPSC, Universit{é} Joseph Fourier/CNRS-IN2P3/INPG, UMR5821, Grenoble,
F-38026, France}

\author{\ \ F.~I.~Olness\,}

\affiliation{Southern Methodist University, Dallas, Texas 75275, USA}

\author{\ \ J.~Y.~Yu\,}

\affiliation{Southern Methodist University, Dallas, Texas 75275, USA}

\author{\ C.~Keppel\,}

\affiliation{Thomas Jefferson National Accelerator Facility, Newport News, Virginia
23602, USA}

\affiliation{Hampton University, Hampton, Virginia, 23668, USA}

\author{\ J.~G.~Morf\'{\i}n\,}

\affiliation{Fermilab, Batavia, Illinois 60510, USA}

\author{\ \ J.F.~Owens\,}

\affiliation{Florida State University, Tallahassee, Florida 32306-4350, USA}

\author{T.~Stavreva\,}

\affiliation{LPSC, Universit{é} Joseph Fourier/CNRS-IN2P3/INPG, UMR5821, Grenoble,
F-38026, France}

\begin{abstract}
We perform a global $\chi^{2}$-analysis of nuclear parton distribution
functions using data from charged current neutrino--nucleus ($\nu A$)
deep-inelastic scattering (DIS), charged-lepton--nucleus ($\ell^{\pm}A$)
DIS, and the Drell-Yan (DY) process. We show that the nuclear corrections
in $\nu A$ DIS are not compatible with the predictions derived from
$\ell^{\pm}A$ DIS and DY data. We quantify this result using a hypothesis-testing
criterion based on the $\chi^{2}$ distribution which we apply to
the total $\chi^{2}$ as well as to the $\chi^{2}$ of the individual
data sets. We find that it is not possible to accommodate the data
from $\nu A$ and $\ell^{\pm}A$ DIS by an acceptable combined fit.
Our result has strong implications for the extraction of both nuclear 
and proton parton distribution functions using combined neutrino and charged-lepton data sets.
\end{abstract}


\maketitle

High statistics neutrino deep-inelastic scattering (DIS) experiments
have generated significant interest in the literature as they provide
crucial information for global fits of parton distribution functions
(PDFs). The neutrino DIS data provide the most stringent constraints
on the strange quark distribution in the proton, and allow for flavor
decomposition of the PDFs which is {\em essential} for precise
predictions of the benchmark gauge boson production processes at the
LHC. Moreover, the neutrino experiments have been used to make precision
tests of the Standard Model (SM). A prominent
example is the extraction of the weak mixing angle $\theta_{W}$ in
a Paschos-Wolfenstein type analysis \citep{Zeller:2001hh}. 
A good knowledge of the neutrino DIS cross sections is also very important 
for long baseline experiments
of the next generation which aim at measuring small parameters of
the mixing matrix such as the mixing angle $\theta_{13}$ and
eventually the CP violating phase $\delta$. %

Because of the weak nature of neutrino interactions the use of heavy nuclear
targets is unavoidable, and this complicates the analysis of the precision
physics discussed above since model-dependent nuclear corrections
must be applied to the data. Our present understanding of the nuclear
corrections is mainly based on charged-lepton--nucleus ($\ell A$)
DIS data. In the early 80s, the European Muon Collaboration (EMC)
\citep{Aubert:1983xm} found that the nucleon structure functions
$F_{2}$ for iron and deuterium differ. This discovery triggered a
vast experimental program to investigate the nuclear modifications
of the ratio $R[F_{2}^{\ell A}] =F_{2}^{\ell A} /(A\, F_{2}^{\ell N})$
for a wide range of nuclear targets with atomic number $A$, see Table~\ref{tab:exp1}.
By now, such modifications have been established in a kinematic range
from relatively small Bjorken $x$ ($x\sim10^{-2}$) to large $x$
($x\sim0.8$) in the deep-inelastic region with squared momentum transfer
$Q^{2}>1\ {\rm GeV}^{2}$. The behavior of the ratio $R[F_{2}^{\ell A}]$
can be divided into four regions: (i) $R>1$ for $x\gtrsim0.8$ (Fermi
motion region), (ii) $R<1$ for $0.25\lesssim x\lesssim0.8$ (EMC
region), (iii) $R>1$ for $0.1\lesssim x\lesssim0.25$ (antishadowing
region), and (iv) $R<1$ for $x\lesssim0.1$ (shadowing region), with
different physics mechanisms explaining the nuclear modifications.
The shadowing suppression at small $x$ occurs due to coherent multiple scattering inside the
nucleus of a $q\bar{q}$ pair coming from the virtual photon \citep{Armesto:2006ph}
with destructive interference of the amplitudes \citep{Brodsky:1989qz}.
The antishadowing region is theoretically less well understood but
might be explained by the same mechanism with constructive interference
of the multiple scattering amplitudes \citep{Brodsky:1989qz} or by
the application of momentum, charge, and/or baryon number sum rules.
Conversely, the modifications at medium and large $x$ are usually
explained by nuclear binding and medium effects and the Fermi motion
of the nucleons \citep{Geesaman:1995yd}.

Instead of trying to address the origin of the nuclear effects, the
data on nuclear structure functions can be analyzed in terms of nuclear
PDFs (NPDFs) which are modified as compared to the free nucleon PDFs.
Relying on factorization theorems in the same spirit as in the free
nucleon case, the advantage of this approach is that the universal
NPDFs can be used to make {\em predictions} for a large variety
of processes in $\ell A$, $\nu A$, $pA$, and $AA$ collisions.
In addition, the nuclear correction factors required for the interpretation
of the neutrino experiments can be calculated in a flexible way, taking
into account the precise observable, the nuclear $A$, and the scale
$Q^{2}$. The factorization assumption in the nuclear environment
is therefore a question of considerable theoretical and practical
importance and global analyses of NPDFs based on $\ell A$ DIS and
fixed target Drell-Yan (DY) data confirm its validity in the presently
explored kinematic range.

\begingroup
\squeezetable
\begin{table}[t]
	\caption{
	The charged-lepton DIS data sets together with DY 
	and with neutrino DIS data sets used in the
	fit. 
	The table details the specific nuclear targets, and the number
	of data points with kinematical cuts ($Q^2>4\ GeV^2$, $W>3.5\ GeV$). References for the data sets
	are cited in Refs.~\citep{Schienbein:2007fs,Schienbein:2009kk}}
\begin{ruledtabular}	
\begin{tabular}{ccclc}
\textbf{\scriptsize ID}{\scriptsize{} } & \textbf{\scriptsize Observable} & \textbf{\scriptsize $A/A' (A)$}  & \textbf{\scriptsize Experiment}{\scriptsize{} } 
& \textbf{\scriptsize $\#$ data}\tabularnewline
\hline 
{\scriptsize 1}  & {\scriptsize $F_{2}^{A}/F_{2}^{A'}$} & {\scriptsize He/D}  & {\scriptsize SLAC-E139}, {\scriptsize NMC-95 re}  & {\scriptsize 15 }\tabularnewline
{\scriptsize 2}  & {\scriptsize $F_{2}^{A}/F_{2}^{A'}$} & {\scriptsize Li/D}  & {\scriptsize NMC-95}  & {\scriptsize 11 }\tabularnewline
{\scriptsize 3}  & {\scriptsize $F_{2}^{A}/F_{2}^{A'}$} & {\scriptsize Be/D}  & {\scriptsize SLAC-E139}  & {\scriptsize 3 }\tabularnewline
{\scriptsize 4}  & {\scriptsize $F_{2}^{A}/F_{2}^{A'}$} & {\scriptsize C/D}  & {\scriptsize EMC-88,90, }{\scriptsize SLAC-E139}  & \tabularnewline
 &  &  & {\scriptsize NMC-95,95 re, }{\scriptsize FNAL-E665-95} & {\scriptsize 38 }\tabularnewline
{\scriptsize 5}  & {\scriptsize $F_{2}^{A}/F_{2}^{A'}$} & {\scriptsize N/D}  & {\scriptsize BCDMS-85}  & {\scriptsize 9 }\tabularnewline
{\scriptsize 6}  & {\scriptsize $F_{2}^{A}/F_{2}^{A'}$} & {\scriptsize Al/D}  & {\scriptsize SLAC-E049,E139}  & {\scriptsize 3 }\tabularnewline
{\scriptsize 7}  & {\scriptsize $F_{2}^{A}/F_{2}^{A'}$} & {\scriptsize Ca/D}  & {\scriptsize EMC-90, }{\scriptsize SLAC-E139}  & \tabularnewline
 &  &  & {\scriptsize NMC-95,re, }{\scriptsize FNAL-E665-95} & {\scriptsize 17 }\tabularnewline
{\scriptsize 8}  & {\scriptsize $F_{2}^{A}/F_{2}^{A'}$} & {\scriptsize Fe/D}  & {\scriptsize BCDMS-85,87}  & \tabularnewline
 &  &  & {\scriptsize SLAC-E049,E139,E140}  & {\scriptsize 24 }\tabularnewline
{\scriptsize 9}  & {\scriptsize $F_{2}^{A}/F_{2}^{A'}$} & {\scriptsize Cu/D}  & {\scriptsize EMC-88,93}  & {\scriptsize 27 }\tabularnewline
{\scriptsize 10} & {\scriptsize $F_{2}^{A}/F_{2}^{A'}$} & {\scriptsize Ag/D}  & {\scriptsize SLAC-E139}  & {\scriptsize 2 }\tabularnewline
{\scriptsize 11} & {\scriptsize $F_{2}^{A}/F_{2}^{A'}$} & {\scriptsize Sn/D}  & {\scriptsize EMC-88}  & {\scriptsize 8 }\tabularnewline
{\scriptsize 12} & {\scriptsize $F_{2}^{A}/F_{2}^{A'}$} & {\scriptsize Xe/D}  & {\scriptsize FNAL-E665-92}  & {\scriptsize 2 }\tabularnewline
{\scriptsize 13} & {\scriptsize $F_{2}^{A}/F_{2}^{A'}$} & {\scriptsize Au/D}  & {\scriptsize SLAC-E139}  & {\scriptsize 3 }\tabularnewline
{\scriptsize 14} & {\scriptsize $F_{2}^{A}/F_{2}^{A'}$} & {\scriptsize Pb/D}  & {\scriptsize FNAL-E665-95}  & {\scriptsize 3 }\tabularnewline
{\scriptsize 15} & {\scriptsize $F_{2}^{A}/F_{2}^{A'}$} & {\scriptsize Be/C}  & {\scriptsize NMC-96}  & {\scriptsize 14 }\tabularnewline
{\scriptsize 16} & {\scriptsize $F_{2}^{A}/F_{2}^{A'}$} & {\scriptsize Al/C}  & {\scriptsize NMC-96}  & {\scriptsize 14 }\tabularnewline
{\scriptsize 17}  & {\scriptsize $F_{2}^{A}/F_{2}^{A'}$}  & {\scriptsize Ca/C}  & {\scriptsize NMC-95,96}  & {\scriptsize 29 }\tabularnewline
{\scriptsize 18}  & {\scriptsize $F_{2}^{A}/F_{2}^{A'}$}  & {\scriptsize Fe/C}  & {\scriptsize NMC-95}  & {\scriptsize 14 }\tabularnewline
{\scriptsize 19}  & {\scriptsize $F_{2}^{A}/F_{2}^{A'}$}  & {\scriptsize Pb/C}  & {\scriptsize NMC-96}  & {\scriptsize 14 }\tabularnewline
{\scriptsize 20}  & {\scriptsize $F_{2}^{A}/F_{2}^{A'}$}  & {\scriptsize C/Li}  & {\scriptsize NMC-95}  & {\scriptsize 7 }\tabularnewline
{\scriptsize 21}  & {\scriptsize $F_{2}^{A}/F_{2}^{A'}$}  & {\scriptsize Ca/Li}  & {\scriptsize NMC-95}  & {\scriptsize 7 }\tabularnewline
{\scriptsize 22}  & {\scriptsize $F_{2}^{A}/F_{2}^{A'}$}  & {\scriptsize He/D}  & {\scriptsize Hermes}  & {\scriptsize 17 }\tabularnewline
{\scriptsize 23}  & {\scriptsize $F_{2}^{A}/F_{2}^{A'}$}  & {\scriptsize Kr/D}  & {\scriptsize Hermes}  & {\scriptsize 12 }\tabularnewline
{\scriptsize 24}  & {\scriptsize $F_{2}^{A}/F_{2}^{A'}$}  & {\scriptsize Sn/C}  & {\scriptsize NMC-96}  & {\scriptsize 111 }\tabularnewline
{\scriptsize 25}  & {\scriptsize $F_{2}^{A}/F_{2}^{A'}$}  & {\scriptsize N/D}  & {\scriptsize Hermes}  & {\scriptsize 19 }\tabularnewline
{\scriptsize 32}  & {\scriptsize $F_{2}^{A}/F_{2}^{A'}$}  & {\scriptsize D}  & {\scriptsize NMC-97}  & {\scriptsize 201 }\tabularnewline
{\scriptsize 26} & {\scriptsize $\sigma_{DY}^{pA}/\sigma_{DY}^{pA'}$} & {\scriptsize C/D}  & {\scriptsize FNAL-E772}  & {\scriptsize 9 }\tabularnewline
{\scriptsize 27} & {\scriptsize $\sigma_{DY}^{pA}/\sigma_{DY}^{pA'}$} & {\scriptsize Ca/D}  & {\scriptsize FNAL-E772}  & {\scriptsize 9 }\tabularnewline
{\scriptsize 28} & {\scriptsize $\sigma_{DY}^{pA}/\sigma_{DY}^{pA'}$} & {\scriptsize Fe/D}  & {\scriptsize FNAL-E772}  & {\scriptsize 9 }\tabularnewline
{\scriptsize 29} & {\scriptsize $\sigma_{DY}^{pA}/\sigma_{DY}^{pA'}$} & {\scriptsize W/D}  & {\scriptsize FNAL-E772}  & {\scriptsize 9 }\tabularnewline
{\scriptsize 30} & {\scriptsize $\sigma_{DY}^{pA}/\sigma_{DY}^{pA'}$} & {\scriptsize Fe/Be}  & {\scriptsize FNAL-E866}  & {\scriptsize 28 }\tabularnewline
{\scriptsize 31} & {\scriptsize $\sigma_{DY}^{pA}/\sigma_{DY}^{pA'}$} & {\scriptsize W/Be}  & {\scriptsize FNAL-E866}  & {\scriptsize 28 }\tabularnewline
&  &  & \textbf{\scriptsize $l^\pm A$ DIS \& DY Total:}{\scriptsize{} } & \textbf{\scriptsize 708}\tabularnewline
\hline
{\scriptsize 33}  &  {\scriptsize $d\sigma^{\nu A}/dx\, dy$} &{\scriptsize Pb}  & {\scriptsize CHORUS $\nu$}  & {\scriptsize 824 }\tabularnewline
{\scriptsize 34}  &  {\scriptsize $d\sigma^{\nu A}/dx\, dy$} &{\scriptsize Pb}  & {\scriptsize CHORUS $\bar{\nu}$}  & {\scriptsize 412 }\tabularnewline
{\scriptsize 35}  &  {\scriptsize $d\sigma^{\nu A}/dx\, dy$} &{\scriptsize Fe}  & {\scriptsize NuTeV $\nu$}  & {\scriptsize 1170 }\tabularnewline
{\scriptsize 36}  &  {\scriptsize $d\sigma^{\nu A}/dx\, dy$} &{\scriptsize Fe}  & {\scriptsize NuTeV $\bar{\nu}$}  & {\scriptsize 966 }\tabularnewline
{\scriptsize 37}  &  {\scriptsize $d\sigma^{\nu A}/dx\, dy$} &{\scriptsize Fe}  & {\scriptsize CCFR di-$\mu$}  & {\scriptsize 44 }\tabularnewline
{\scriptsize 38}  &  {\scriptsize $d\sigma^{\nu A}/dx\, dy$} &{\scriptsize Fe}  & {\scriptsize NuTeV di-$\mu$}  & {\scriptsize 44 }\tabularnewline
{\scriptsize 39}  &  {\scriptsize $d\sigma^{\nu A}/dx\, dy$} &{\scriptsize Fe}  & {\scriptsize CCFR di-$\mu$}  & {\scriptsize 44 }\tabularnewline
{\scriptsize 40}  &  {\scriptsize $d\sigma^{\nu A}/dx\, dy$} &{\scriptsize Fe}  & {\scriptsize NuTeV di-$\mu$}  & {\scriptsize 42 }\tabularnewline
 &  &  & \textbf{\scriptsize $\nu A$ Total:}{\scriptsize{} } & \textbf{\scriptsize 3134}\tabularnewline
\end{tabular}
\end{ruledtabular}	 
{\scriptsize \label{tab:exp1}} 
\end{table}
\endgroup
\begingroup
\squeezetable
\begin{table}[t]
	\caption{Summary table of a family of compromise fits. \label{tab:compr} }
\begin{ruledtabular}	
\begin{tabular}{lccccc}
$w$  & $l^{\pm}A$  & $\chi^{2}$ (/pt)  & $\nu A$  & $\chi^{2}$ (/pt)  & total $\chi^{2}$(/pt)\tabularnewline
\hline 
$0$  & 708  & 638 (0.90)  & -  & -  & 638 (0.90) \tabularnewline
$1/7$  & 708  & 645 (0.91)  & 3134  & 4710 (1.50)  & 5355 (1.39) \tabularnewline
$1/2$  & 708  & 680 (0.96)  & 3134  & 4405 (1.40)  & 5085 (1.32) \tabularnewline
$1$  & 708  & 736 (1.04)  & 3134  & 4277 (1.36)  & 5014 (1.30) \tabularnewline
$\infty$  & -  & -  & 3134  & 4192 (1.33)  & 4192 (1.33) \tabularnewline
\end{tabular}
\end{ruledtabular}
\end{table}
\endgroup

However, in a recent analysis \citep{Schienbein:2007fs} of $\nu Fe$
DIS data from the NuTeV collaboration we found that the nuclear correction
factors are surprisingly different from the predictions based on the
$\ell^{\pm}Fe$ charged-lepton results with important implications
for global analyses of proton PDFs. This 
is not completely
unexpected since the structure functions in charged current (CC) neutrino
DIS and neutral current (NC) electron or muon DIS are distinct observables
with different parton model expressions. From this perspective it
is clear that the nuclear correction factors will not be exactly the
same even for a {\em universal} set of NPDFs. Note also that some
models in the literature predict differences between reactions in
CC and NC DIS \citep{Brodsky:2004qa}. What is, however, unexpected
is the degree to which the $R$ factors differ between the structure
functions $F_{2}^{\nu Fe}$ and $F_{2}^{\ell Fe}$. In particular
the lack of evidence for shadowing in neutrino scattering down to
$x\sim0.02$ is quite surprising.

The study in Ref.~\citep{Schienbein:2007fs} left open the question, whether
the neutrino DIS data could be reconciled with the charged-lepton
DIS data by a better flavor separation of the NPDFs. In this letter,
we address this question in the $A$-dependent framework of Ref.~\citep{Schienbein:2009kk}
by performing a global $\chi^{2}$-analysis of the combined data from
$\nu A$ DIS, $\ell A$ DIS and the DY process listed in Table~\ref{tab:exp1}.

When combining neutrino and charged-lepton+DY data into a compromise
fit, we introduce a weight parameter $w$ into the $\chi^{2}$ via:
\begin{equation}
\chi^{2}=\sum_{l^{\pm}A\ {\rm data}}\chi_{i}^{2}\ +\!\!\sum_{\nu A\ {\rm data}}w\chi_{i}^{2}\quad.\label{eq:chi2}
\end{equation}
The $w$ factor allows us to adjust for the different number of points
in the separate data sets, and provides a parameter that 
interpolates between the $\nu A$ and the $\ell^{\pm}A$+DY data.
We should stress that the $\chi^{2}$ cited in Table~\ref{tab:compr} 
and also in the text is the standard $\chi^{2}$;
Eq.~(\ref{eq:chi2}) is only used internally in the fitting procedure.
We construct a set of compromise fits with weights $w=\{0,\frac{1}{7},\frac{1}{2},1,\infty\}$
and study the dependence of the result on this weight. The fit to only
neutrino data, denoted $w=\infty$ in Table~\ref{tab:compr}, is compatible
with the results in \citep{Schienbein:2007fs}. Similarly, the fit
to only charged-lepton+DY data, denoted $w=0$, agrees well with the
analysis in \citep{Schienbein:2009kk}.

\begin{figure}[b]
\includegraphics[width=0.45\textwidth]{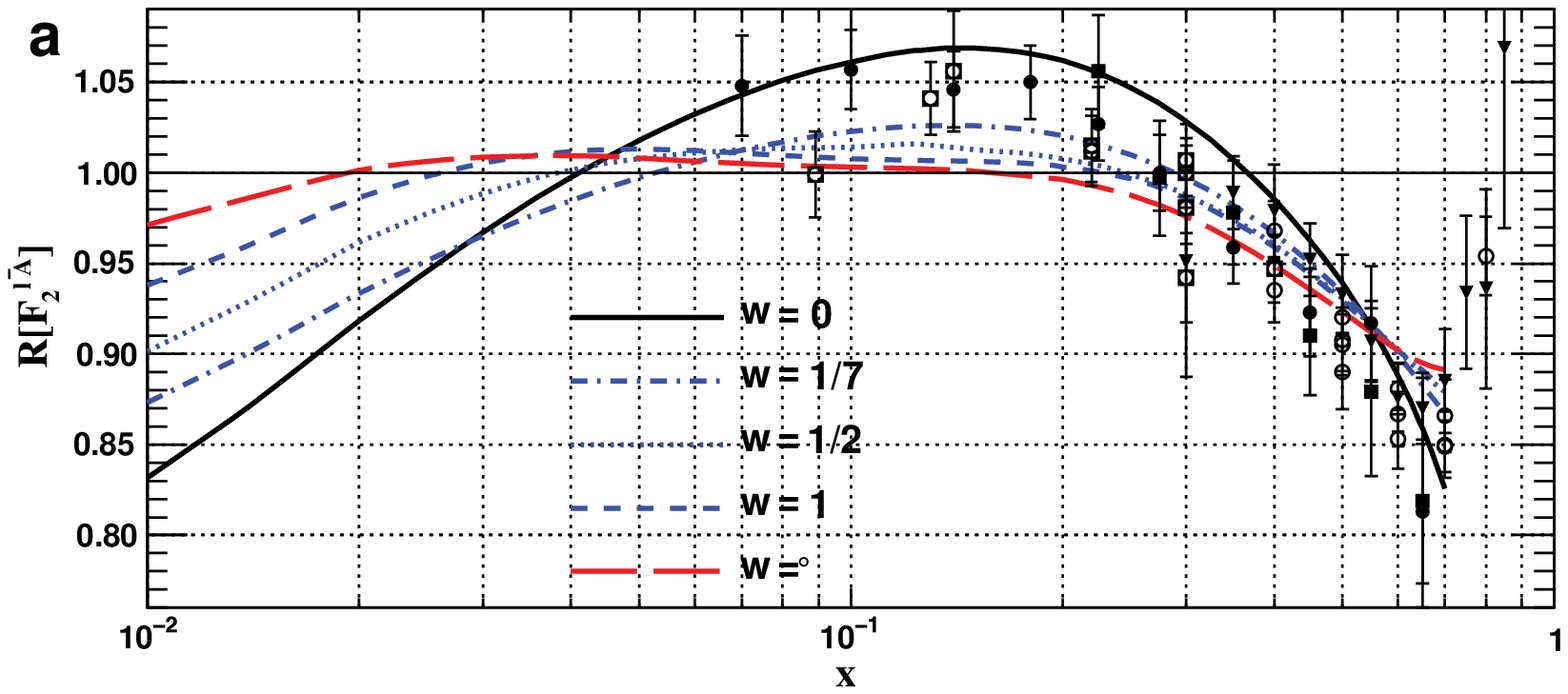}\hspace{1cm}
\includegraphics[width=0.45\textwidth]{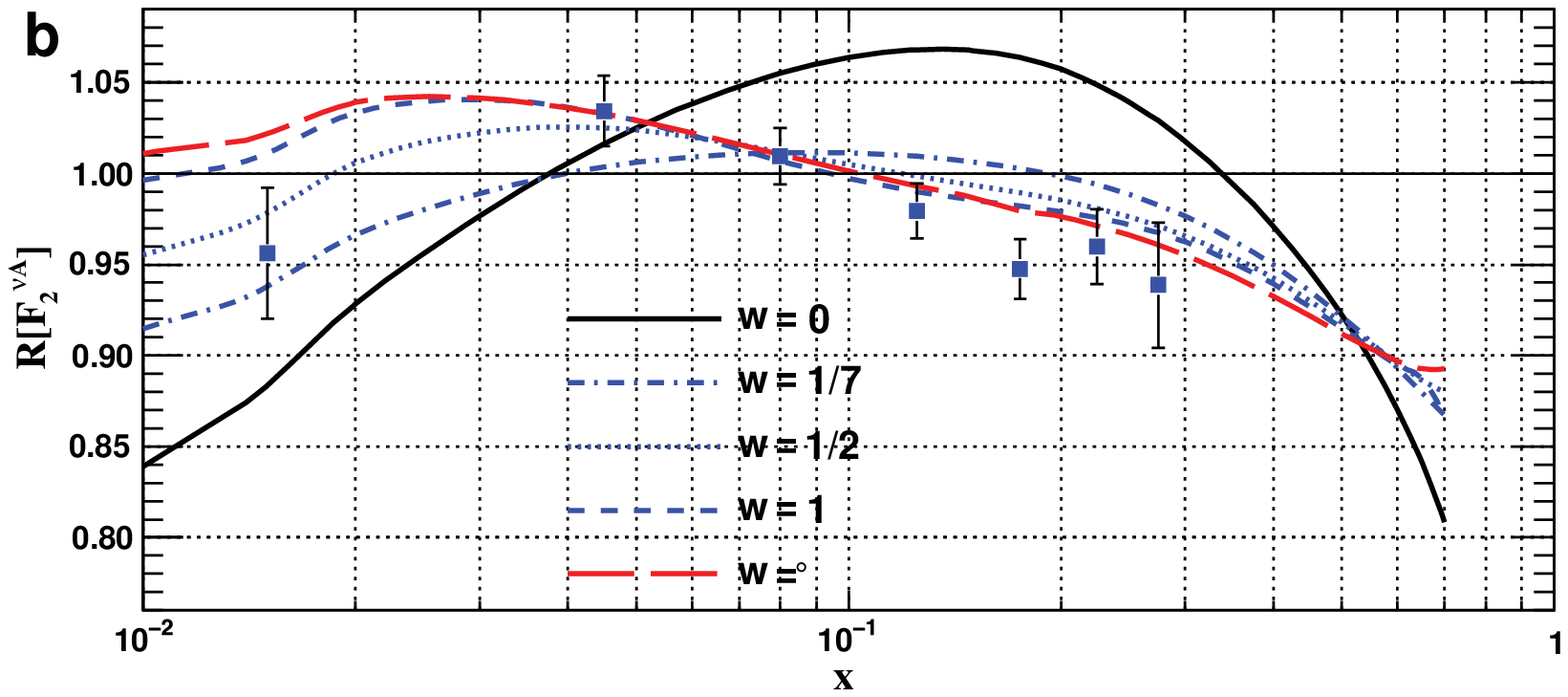}
\caption{ Predictions from the compromise fits for the nuclear correction factors
$R[F_{2}^{\ell Fe}]\simeq F_{2}^{\ell Fe}/F_{2}^{\ell N}$ (a)
and $R[F_{2}^{\nu Fe}]\simeq F_{2}^{\nu Fe}/F_{2}^{\nu N}$ (b)
as a function of $x$ for $Q^{2}=5\,{\rm GeV}^{2}$. The data points
displayed in (a) are from BCDMS and SLAC experiments (for references 
see \citep{Schienbein:2009kk})
and those displayed in (b) come from the NuTeV experiment \citep{Tzanov:2005kr}.
\label{fig:F2compare}}
\end{figure}

We first compute the nuclear correction factors $R[F_{2}^{\ell Fe}]\simeq F_{2}^{\ell Fe}/F_{2}^{\ell N}$
and $R[F_{2}^{\nu Fe}]\simeq F_{2}^{\nu Fe}/F_{2}^{\nu N}$ %
in the QCD parton model at next-to-leading
order employing the NPDF fits in Table~\ref{tab:compr} for the numerator
and free nucleon PDFs for the denominator.
\footnote{We allow each individual partonic flavor to have different nuclear
corrections, and the details are outlined in
Refs.~\citep{Schienbein:2007fs,Schienbein:2009kk}.
While we focus on $F_2$ , we can consider other observables such as
$\{ F_1,F_3,d\sigma\} $ in a similar manner.} 
The $x$-dependence of
$R[F_{2}^{\ell Fe}]$ and $R[F_{2}^{\nu Fe}]$ is shown in Fig.~\ref{fig:F2compare}
a) and b), respectively, at $Q^{2}=5\ {\rm GeV}^{2}$. Similar results
hold at $Q^{2}=20\ {\rm GeV}^{2}$ which we do not present here. We
observe that the fit to only $\ell A$ DIS+DY data ($w=0$) well describes
the SLAC and BCDMS points
in Fig.~\ref{fig:F2compare} a). The same is true for the fit to
only $\nu A$ DIS data ($w=\infty$) which is compatible with the
results from the NuTeV experiment \citep{Tzanov:2005kr}
exemplified in Fig.~\ref{fig:F2compare} b). However, comparing the
results obtained with the $w=0$ and the $w=\infty$ fits one can
see that they predict considerably different $x$-shapes.
\newline %
The fits with weights $w=\{\frac{1}{7},\frac{1}{2},1\}$ interpolate between these two
incompatible solutions. As can be seen in Fig.~\ref{fig:F2compare}
a) and b), with increasing weight the description of the $\ell Fe$
data is worsened in favor of a better agreement with the $\nu Fe$
points. This trend clearly demonstrates that the $\ell Fe$ and the
$\nu Fe$ data pull in opposite directions. We identify the fits with
$w=1/2$ or $w=1$ as the best candidates for a possible compromise.
\begin{figure}
\includegraphics[clip,width=0.45\textwidth]{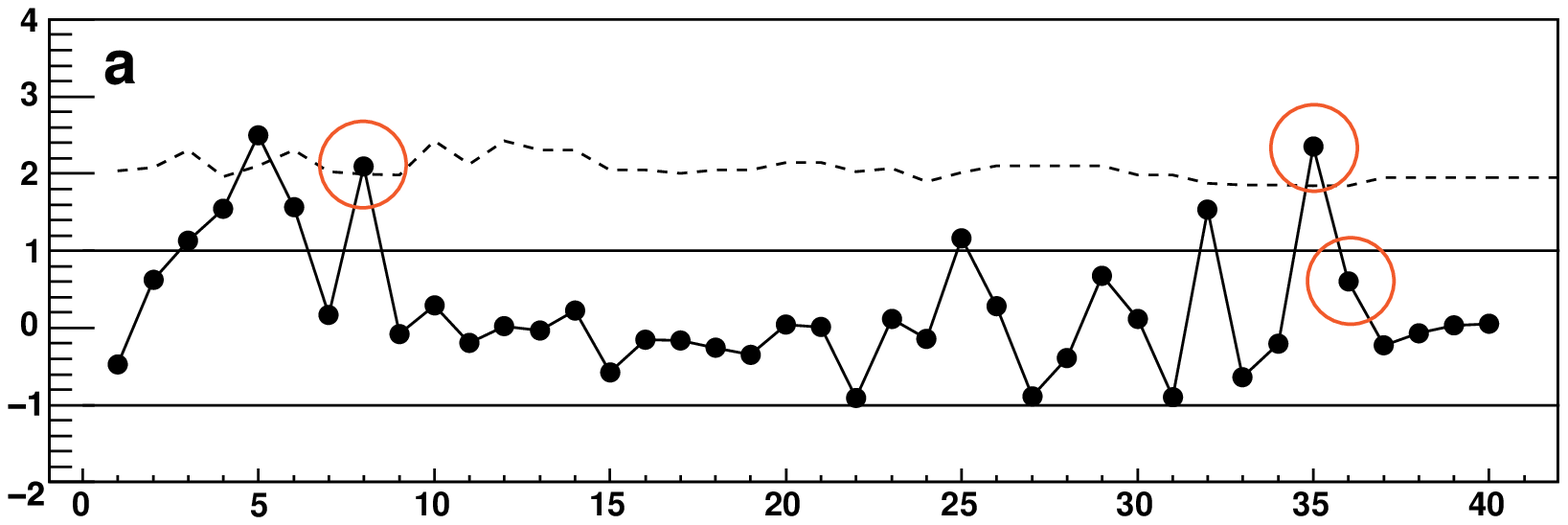}\hspace{1cm}
\includegraphics[clip,width=0.45\textwidth]{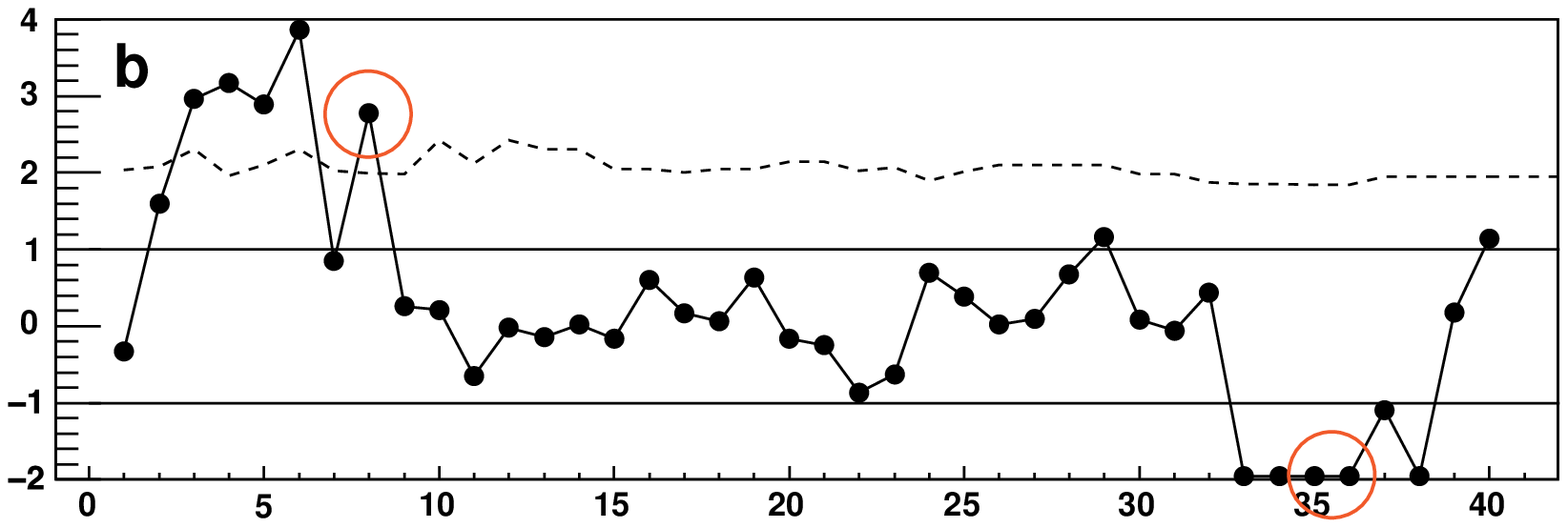}
\caption{$\Delta\chi^{2}/\Delta C_{90}$ as defined in Eq.~(\ref{eq:deltachisq})
for the 40 individual data sets. Results are shown for the $w=\frac{1}{2}$-fit
(a) and the fit `Ucor5' (b) with $w=1$. The solid and dashed lines indicate
the $90\%$ and $99\%$ confidence limits. The highlighted data sets
correspond to DIS $\ell^{\pm}Fe$ (ID=8), $\nu Fe$ (ID=35),
and $\bar{\nu}Fe$ (ID=36). \label{fig:delchi2}}
\end{figure}
To be able to decisively accept or reject the compromise fits, we
apply a statistical goodness-of-fit criterion \citep{Stump:2001gu,Martin:2009short,Eskola:2009uj}
based on the probability distribution for the $\chi^{2}$ given that
the fit has $N$ degrees of freedom: \begin{equation}
P(\chi^{2},N)=\frac{(\chi^{2})^{N/2-1}e^{-\chi^{2}/2}}{2^{N/2}\Gamma(N/2)}\,.\label{eq:chi2dist}\end{equation}
 This allows us to define the percentiles $\xi_{p}$ via $\int_{0}^{\xi_{p}}P(\chi^{2},N)d\chi^{2}=p\%$
where $p=\{50,90,99\}$. Here, $\xi_{50}$ serves as an estimate of
the mean of the $\chi^{2}$ distribution and $\xi_{90}$, for example,
gives us the value where there is only a 10\% probability that a fit
with $\chi^{2}>\xi_{90}$ genuinely describes the given set of data.
In a global PDF fit, the best fit $\chi^{2}$ value often deviates
from the mean value because the data come from different possibly
incompatible experiments having unidentified, unknown errors which
are not accounted for in the experimental systematic errors. For this
reason we rescale the $\xi_{90}$ and $\xi_{99}$ percentiles relative
to the best fit $\chi_{0}^{2}$ \citep{Stump:2001gu} to define $C_{90}=\chi_{0}^{2}(\xi_{90}/\xi_{50})$and
$C_{99}=\chi_{0}^{2}(\xi_{99}/\xi_{50})$. This defines our criterion:
a fit with a given $\chi^{2}$ is compatible with the best fit with
$\chi_{0}^{2}$ at $90\%$ ($99\%$) confidence if $\chi^{2}<C_{90}$
($\chi^{2}<C_{99}$). We apply it to both the total $\chi^{2}$ and
the $\chi^{2}$ of the individual data sets.
\newline %
For the $\ell A$ DIS+DY data we use the fit with $w=0$ as benchmark
with $\chi_{0}^{2}=638$ and $N=677$ degrees of freedom (for $708$
data points and $31$ free parameters). The upper limits on the $\chi^{2}$
at $90\%$ and $99\%$ confidence level (C.L.) are then $C_{90}^{l^{\pm}A}=684$
and $C_{99}^{l^{\pm}A}=722$. The benchmark fit for the $\nu A$
DIS data ($w=\infty$) uses $3134$ data points with $33$ free parameters
resulting in $N=3101$ and one finds $C_{90}^{\nu A}=4330$ and $C_{99}^{\nu A}=4445$.
We see that none of the compromise fits satisfies both limits at the
$90\%$ C.L. which is usually used in global analyses of PDFs to define
the uncertainty bands. At the $99\%$ C.L., there are two fits ($w=1/2$,
$w=1$) which are below the $C_{99}^{\nu A}$ limit. However, only
the $w=1/2$ fit satisfies the corresponding constraint from the charged-lepton
benchmark fit.
\newline %
We now apply our criterion also to the individual data sets with IDs
between 1 and 40 in Table~\ref{tab:exp1}. For the $\ell A$ DIS+DY
data (ID=[1,31]) we determine the 31 $C_{90}$ ($C_{99}$) limits
by using the individual $\chi_{i}^{2}$ of the $w=0$ fit as $\chi_{0,i}^{2}$.
For the $\nu A$ DIS data (ID=[33,40])
we proceed in a similar manner using the individual $\chi_{i}^{2}$
of the $w=\infty$ fit. The results of this detailed analysis are
depicted in Fig.~\ref{fig:delchi2}, where we show the quantity \begin{equation}
\frac{\Delta\chi^{2}}{\Delta C_{90}}=\frac{\chi_{i}^{2}-\chi_{0,i}^{2}}{C_{90,i}-\chi_{0,i}^{2}}\quad(i=1,\ldots,40)\,,\label{eq:deltachisq}\end{equation}
where $\chi_{i}^{2}$ represents the $\chi^{2}$-value of the $i$'th
data set. In cases where $\chi_{i}^{2}>C_{90,i}$ the fit is not compatible
with the best fit at the $90\%$ level and $\Delta\chi^{2}/\Delta C_{90}>1$.
The exact $90\%$ C.L. limit is shown as a constant solid line and
the dotted line represents the $99\%$ confidence limit. The local
application of the $\chi^{2}$ hypothesis-testing criterion reveals
that even the compromise fit with weight $w=\frac{1}{2}$ which was
considered acceptable at the 99\% C.L. when looking at the nuclear
correction factors and at the global change in $\chi^{2}$, cannot
be accepted as a compromise solution as both the charged-lepton and
neutrino DIS data on iron exceed the $99\%$ limit.
\newline %
In conclusion, the tension between the $\ell^{\pm}Fe$ and $\nu Fe$
data sets leaves us with no possible compromise fit when investigating
the results in detail, not even when using the $99\%$ percentile
as the limit as opposed to the more restrictive $90\%$ limit which
is usually used to construct the error PDFs. This detailed analysis
confirms the preliminary conclusions of Refs.~\citep{Schienbein:2007fs,Schienbein:2009kk}
that there is no possible compromise fit which adequately describes 
the neutrino DIS data along with the 
charged-lepton data.
\newline %
At face value, this conclusion differs from some results in the literature
which argue the $\nu A$ and $\ell^{\pm}A$ data are in accord \citep{Paukkunen:2010hb}.
Here, we believe an essential element in our analysis is the use of
the correlated systematic errors of the $\nu A$ data. To highlight
this point, we now repeat our analysis, \emph{but} we combine the
statistical and all 
systematic errors in quadrature (thereby neglecting
the information contained in the correlation matrix) for $\nu A$
data for the $w=1$ fit with $Q^{2}>4\, GeV$ (as before); we denote
this the {}``Ucor4'' fit, and we obtain $\chi^{2}/pt$ of 1.14 for
$\ell^{\pm}A$ and 1.00 for $\nu A$. We also use a $Q^{2}>5\, GeV$
fit (denoted {}``Ucor5'') to mimic the cuts of Ref.~\citep{Paukkunen:2010hb};
here we obtain $\chi^{2}/pt$ of 
1.14 for $\ell^{\pm}A$ and
0.96 for $\nu A$. 
\newline %
If we examine the total $\chi^{2}$ values, we find the $\chi^{2}/dof\sim1$,
and might be tempted to conclude we are able to fit both the $\nu A$
and $\ell^{\pm}A$ data simultaneously. However, if we look at individual
data sets and apply our hypothesis-testing criteria, the picture is
quite different. Figure~\ref{fig:delchi2} (b) displays the results
for the Ucor5 fit. The higher $Q^{2}$ cut of the Ucor5 fit removes
some of the very precise NuTeV data at small-$x$, thus resulting
in an improved $\chi^{2}$ compared to Ucor4. Nevertheless, many of
the $\ell^{\pm}A$ data sets (ID=3,4,5,6,8) still lie outside the
$99\%$ CL percentile. Thus, we still conclude that there is no compromise
fit for the $\nu A$ and $\ell^{\pm}A$ data even if we relax the
constraints by using uncorrelated errors.
\newline %
Consequently, the nuclear correction factor for the 
neutrino DIS data are indeed {\em incompatible}
with that of the charged-lepton DIS and DY data, and this result depends crucially
on the use of the precision correlated errors of the neutrino data.
This result has important implications for both nuclear and proton
PDFs. 
If we do not know the appropriate nuclear correction to relate
different nuclear targets, our ability to extract PDFs is
limited. 
For example, the CTEQ6.6 analysis \citep{Nadolsky:2008zw}
sidesteps these issues by removing most of the $\nu A$ data from the
fit; however, they retain the NuTeV dimuon data since this data is
critical to constraining the strange quark PDF. This underscores the
importance of the $\nu A$ data for flavor differentiation. 
\newline %
Although the NuTeV data provide the tightest constraints due to their
statistics, we note that this issue cannot be tied to a single data
set. For example, we find that NuTeV is generally compatible with
CCFR and CDHSW%
\footnote{For $x<0.4$, NuTeV is compatible with both CCFR and CDHSW data; for
larger $x$, NuTeV agrees with CDHSW, and the difference with CCFR
has been reconciled.%
}. The CHORUS $\nu Pb$ and $\bar{\nu}Pb$ data have larger uncertainties,
so they can be compatible with both the $\ell^{\pm}A$ data and the
NuTeV $\nu Fe$ data because the $\Delta\chi^{2}/\Delta C_{90}<1$
for all weights. Compared to the theory predictions, NuTeV agrees
well in the central $x$ region, but exhibits differences both for
low $x$ at low $Q^{2}$, and also for very high $x$ ($x\sim 0.65$). 
\newline %
We have demonstrated that the $\nu A$ and $\ell^{\pm}A$ data prefer
different nuclear correction factors, and that there is no single
{}``compromise'' result that will simultaneously satisfy both data
sets. While we have focused on the phenomenological aspects for the
present study, this result has strong implications for the extraction
of both nuclear and proton PDFs using combined neutrino and charged-lepton
data sets. Possibilities include unexpectedly large higher-twist effects,
or even nonuniversal nuclear effects; we leave such questions for
a future study.

\bibliographystyle{apsrev4-1}
\bibliography{npdf}

\end{document}